\documentclass[aps,pre,floats,floatfix,twocolumn,showpacs,superscriptaddress]{revtex4-1}

\usepackage{latexsym,amsmath}
\usepackage{amsbsy}
\usepackage[pdftex]{graphicx}
\usepackage{amssymb}
\usepackage{epstopdf}
\usepackage[usenames]{xcolor}

\begin{document}

\title{Escaping the avalanche collapse in self-similar multiplexes}

\author{M. \'Angeles Serrano}
\affiliation{Departament de F\'isica Fonamental, Universitat de Barcelona, Mart\'{\i} i Franqu\`es 1, 08028 Barcelona, Spain}

\author{\v{L}ubo\v{s} Buzna}
\affiliation{Departament de F\'isica Fonamental, Universitat de Barcelona, Mart\'{\i} i Franqu\`es 1, 08028 Barcelona, Spain}
\affiliation{University of \v{Z}ilina, Univerzitn\'{a} 8215/1, 01026 \v{Z}ilina, Slovakia}

\author{Mari\'an Bogu\~n\'a}
\affiliation{Departament de F\'isica Fonamental, Universitat de Barcelona, Mart\'{\i} i Franqu\`es 1, 08028 Barcelona, Spain}

\date{\today}
            
\begin{abstract}
We deduce and discuss the implications of self-similarity for the stability in terms of robustness to failure of multiplexes, depending on interlayer degree correlations. First, we define self-similarity of multiplexes and we illustrate the concept in practice using the configuration model ensemble. Circumscribing robustness to survival of the mutually percolated state, we find a new explanation based on self-similarity both for the observed fragility of interconnected systems of networks and for their robustness to failure when interlayer degree correlations are present. Extending the self-similarity arguments, we show that interlayer degree correlations can change completely the stability properties of self-similar multiplexes, so that they can even recover a zero percolation threshold and a continuous transition in the thermodynamic limit, qualitatively exhibiting thus the ordinary stability attributes of noninteracting networks. We confirm these results with numerical simulations.
\end{abstract}

\maketitle

\section{Introduction}
Self-similarity is defined in a wide sense as the property of some systems to be, either exactly or statistically, similar to a part of themselves. This property is found in certain geometric objects that are intrinsically embedded in metric spaces, so that distance in the metric space gives a natural standard of measurement to uncover similar patterns at different observation scales~\cite{Mandelbrot:1967}. In complex networks, the definition of self-similarity is not obvious since many networks are not explicitly embedded in any physical geometry and the only available metric is the one induced by the collection of shortest path lengths between nodes. This metric has, in fact, been used to measure the fractal and self-similar properties of complex networks~\cite{Song:2005uq,Goh:2006ee}. However, the small-world property typically found in real complex networks strongly limits the range of scales where such properties can be observed.

In the absence of a natural geometry, the main problem in the definition of self-similarity stems from the fact that there is, a priori, no way to decide what is the ``part'' of the system that should be compared to (and look alike) the ``whole''. In this sense, self-similarity is not an intrinsic property of the system but it is directly related to the specific procedure to identify the appropriate subsystem. In previous work on single networks, self-similarity was properly defined on the basis of a nested hierarchy of subgraphs and proved for general classes of models. These include random scale-free models with and without underlying metric spaces and models of growing networks~\cite{mariangels,Serrano:2011kq}. 
Interestingly, metric network models are able to provide a plausible explanation for key topological properties observed in real networks~\cite{Boguna:2009uz,Serrano:2012we,Papadopoulos:2012uq}, including scale-free degree distributions, high levels of clustering, the small world property, and self-similarity.

Self-similarity has important implications in the global structure of networks and, in particular, in their vulnerability to failures of their constituents. For instance, self-similarity alone ---independently of the divergence of the second moment of the degree distribution--- explains the absence of a percolation threshold in random scale-free networks, with a proof that avoids the usual locally tree-like and other limiting assumptions~\cite{Serrano:2011kq}. Moreover, the same proof applies to ensembles of graphs with highly non-trivial topologies as long as they belong to the same self-similarity class. In~\cite{Serrano:2011kq}, the absence of a percolation threshold was also proved and numerically confirmed in ensembles of random networks embedded in metric spaces with strong clustering and in ensembles of growing networks with bounded topological fluctuations.

In this work, we extend the concept of self-similarity to multiplexes ---defined as networks of nodes interconnected with different classes of links, each class named a layer. Out of the many different self-similar ensembles in single networks, we chose for simplicity the configuration model and generalize it to multiplexes in order to state explicitly the definition and significance of self-similarity in such structures. In particular, we study the implications of self-similarity for the stability and robustness to failure of multiplexes with and without interlayer degree correlations. Circumscribing robustness to survival of the mutually percolated state~\cite{BPPSH10,Son:2012}, we find a new explanation based on self-similarity both for the observed fragility of uncorrelated scale-free systems of networks~\cite{BPPSH10,Baxter:2012} and for their robustness to failure when correlations are present~\cite{Buldyrev:2011fk,Reis:2014}. We find that interlayer degree correlations can change completely the stability of self-similar scale-free multiplexes, which can recover a zero percolation threshold and a continuous transition in the thermodynamic limit qualitatively exhibiting so the ordinary stability properties of single scale-free networks. 

The paper is organized as follows. In section~\ref{sec:II}, we review the definition of self-similarity in single-layered networks and extend it to multiplexes. In section~\ref{sec:III}, we discuss the self-similarity properties of the canonical configuration model generalized to multiplexes, both with and without interlayer degree correlations. In section~\ref{sec:IV}, we use this model to deduce and discuss the implications of self-similarity on mutual percolation and check our predictions against numerical simulations. Finally, we conclude in section~\ref{sec:V}.

\section{Self-similar ensembles}
\label{sec:II}

In the next section, we first review our findings on this topic in the case of single networks and, then, extend them to the case of multiplexes.

\subsection{One-layered self-similar ensembles}
Let ${\cal G}(\{ \alpha \})$ be an ensemble of sparse graphs in the thermodynamic limit, where $\{ \alpha \}$ is the set of model parameters. For example, in the case of the Erd\"os-R\'enyi model~\cite{Erdos:1959yo,Erdos:1960dl} the set $\{ \alpha \}$ is just the average degree $\langle k \rangle$. Consider a transformation rule $T$ that for each graph $G \in {\cal G}(\{ \alpha \})$ selects one of $G$'s subgraphs. Denote the ensemble of these subgraphs by ${\cal G}_T(\{ \alpha \})$. The ensemble ${\cal G}(\{ \alpha \})$ is called self-similar with respect to $T$ if the transformed ensemble is the same as the original one except for some transformation of the model parameters, that is,
\begin{equation}
{\cal G}_T(\{ \alpha \})={\cal G}(\{ \alpha_T \}),
\label{eq:ss-def}
\end{equation}
where $\{ \alpha_T \}$ are the ensemble parameters after the filtering process. This definition does not assume anything about the transformation rule $T$ and, in fact, the same ensemble can be self-similar under different rules. As a simple example consider the Erd\"os-R\'enyi model with $N\gg1$ nodes and connection probability among pairs of nodes $p=\langle k \rangle/N$. Now consider the transformation rule that selects $N_T$ nodes uniformly at random out of the original $N$ nodes, along with their connections. It is easy to see that such subgraph belongs to the Erd\"os-R\'enyi ensemble but with an average degree 
\begin{equation}
\langle k \rangle_T=\frac{N_T}{N} \langle k \rangle.
\end{equation}
Note that the average degree of subgraphs generated with this procedure is smaller than the average degree of the original network.
\begin{figure}[t]
\centerline{\includegraphics[width=\linewidth]{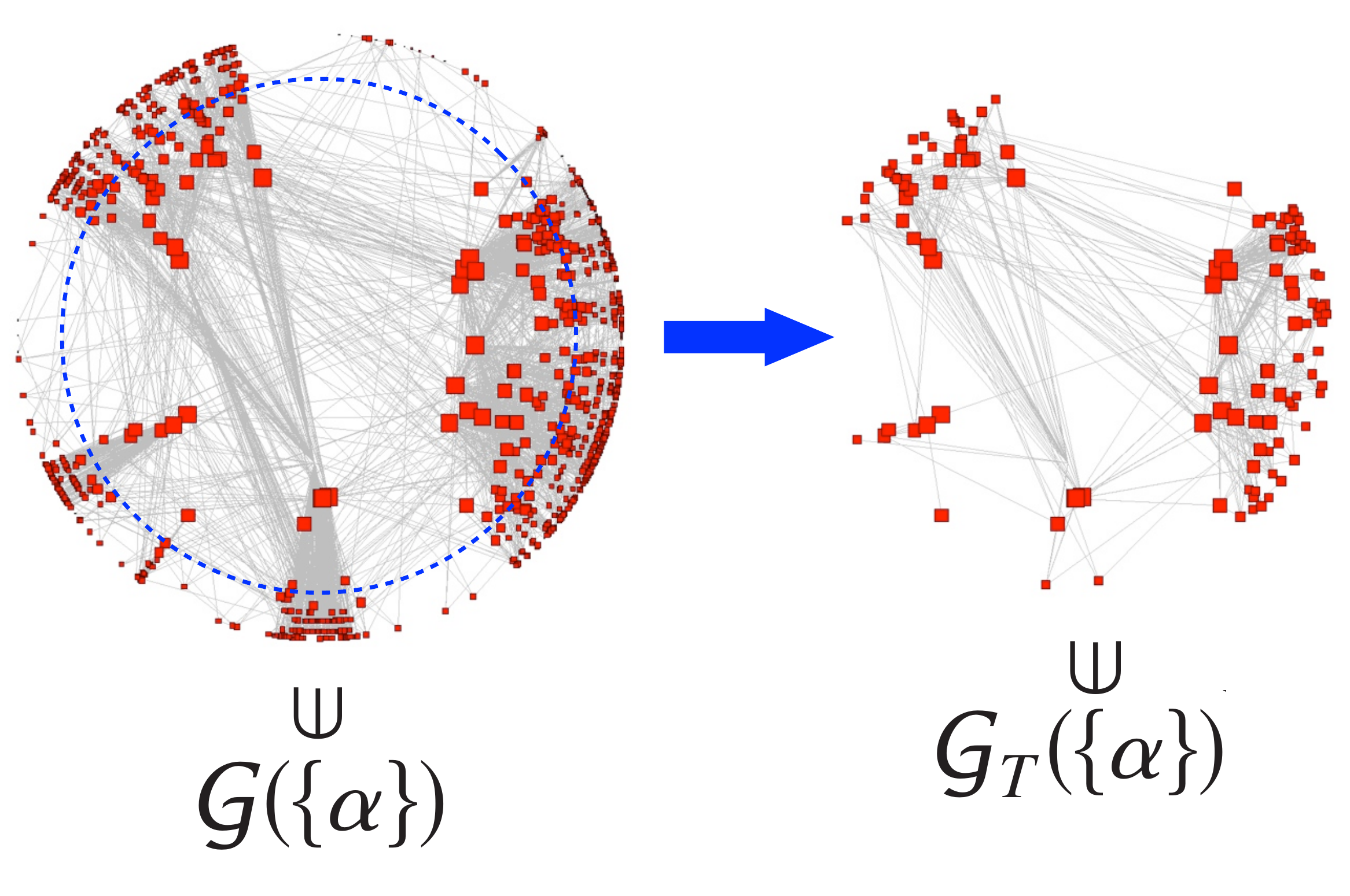}}
\caption{Illustration of a self-similar ensemble of graphs embedded into a metric space, a circle of radius $R\sim N$, under a transformation that removes nodes with degrees below a certain threshold~\cite{mariangels}. In this visualization, each node is given a radial coordinate inversely proportional to its degree so that we obtain the desired subgraph by removing all nodes outside the blue dashed circle.\label{fig:1}} 
\end{figure}

In the ensembles studied in~\cite{mariangels,Serrano:2011kq} ---including the standard configuration model with scale-free degree distributions and zero clustering, scale-free networks with finite clustering and metric structure, and non-equilibrium networks, like generic growing network models---, the only model parameter that changes after the transformation is the average degree of the subgraph, $\langle k \rangle_T$. Typically, this average is a monotonic function of the ratio between the size of the original network $N$ and the size of the subgraph $N_T$, that is, 
\begin{equation}
\langle k \rangle_T=f\left(\frac{N}{N_T}\right) \langle k \rangle.
\end{equation}
In this case, the sign of its derivative determines the class of self-similarity of the model and, in turn, the structural properties of the entire network. For instance, when $f(x)$ is a monotonic increasing function, any graph of the ensemble contains subgraphs with an arbitrary large average degree within the subgraph. This is the case of the configuration model with an scale-free degree distribution with exponent $2<\gamma<3$ and, remarkably, of many real-world networked systems~\cite{mariangels}. This simple property, together with the fact that these subgraphs belong to the same ensemble, imply a zero percolation threshold in the thermodynamic limit~\cite{Serrano:2011kq}, even if $\gamma>>3$. Remarkably, this is a consequence of self-similarity alone and not of the divergence of the second moment of the degree distribution. The proof in~\cite{Serrano:2011kq}  represents a powerful alternative to typical techniques applied to the study of percolation in complex networks, since it avoids the usual locally tree-like and other limiting assumptions. 

\subsection{Self-similar multiplexes}
Formally, self-similarity of random multiplexes can be defined as for single networks. As in Eq.~({\ref{eq:ss-def}}), let ${\cal M}(\{ {\bf \alpha} \})$ be a multiplex ensemble of sparse graphs in the thermodynamic limit, where $\{ {\bf \alpha} \}$ is the set of model parameters, now including sets of model parameters for each layer. Consider a transformation rule $T$ that for each multiplex $M \in {\cal M}(\{ {\bf \alpha} \})$ selects one of $M$'s subgraphs. This transformation rule selects nodes in the multiplex according to specific conditions imposed on each layer. Denote the ensemble of subgraphs by ${\cal M}_T(\{ \alpha \})$. The ensemble ${\cal M}(\{ \alpha \})$ is called self-similar with respect to the transformation rule $T$ if the transformed multiplex ensemble is the same as the original one except for some transformation of the model parameters, that is,
\begin{equation}\label{eq:Mss-def}
{\cal M}_T(\{ {\bf \alpha} \})={\cal M}(\{ {\bf \alpha}_T \}).
\end{equation}
To get insights on the nature and consequences of self-similarity in multiplexes, hereafter we focus on the soft version of the configuration model, the simplest self-similar ensemble with a non-trivial degree distribution~\cite{Serrano:2011kq}. Nevertheless, the generalization to other ensembles is straightforward.

\section{The soft configuration model}
\label{sec:III}
The configuration model is defined as the maximally random ensemble of graphs with a given degree sequence, that is, a predefined degree assigned to each single node of the network~\cite{Bollobas:1980te,Bekessy:1972bp,Bender:1978id}. The soft configuration model (SCM) is very similar to the original one except that, in this case, nodes are given their expected degrees and not their actual degrees~\cite{Chung:2002fl,Boguna:2003um}. This makes the model more appropriate to deal with structural topological correlations that are unavoidable when the degree distribution is broadly distributed~\cite{Boguna:2004eh,Serrano:2007nl}.

In the particular case of scale-free networks, graphs are generated by assigning to each of the $N$ nodes a hidden variable $\kappa$ drawn from a power-law probability density $\rho(\kappa)=(\gamma-1) \kappa_0^{\gamma-1} \kappa^{-\gamma}$, $\kappa \ge \kappa_0$. 
 Nodes with expected degrees $\kappa$ and $\kappa'$ are then connected with probability $r(\kappa,\kappa') \equiv r(\mu \kappa \kappa')$, where function $r(x) \le 1$ is an arbitrary function with $r(0)=0$ and $r'(0) \neq 0$. Constant $\mu$ fixes the average degree $\langle k \rangle$ through the relation
\begin{equation}
\mu=\frac{\langle k \rangle}{N r'(0) \langle \kappa \rangle^2}=\frac{\langle k \rangle}{N r'(0) \kappa_0^2} \left(\frac{\gamma-2}{\gamma-1} \right)^2.
\label{mu}
\end{equation}
With this choice, it is easy to see that the average degree of a node with hidden variable $\kappa$ is proportional to $\kappa$, so that the degree distribution scales as well as a power law with exponent $\gamma$~\cite{Boguna:2004eh}. When function $r(x)$ is chosen to be
\begin{equation}
r(x)=\frac{1}{1+1/x},
\label{r_typeI}
\end{equation}
the model produces maximally random graphs with a given expected degree sequence~\cite{Park:2003qy,Garlaschelli:2008fd,Anand:2009am}. Random graphs with arbitrary structural correlations can be generated as well by choosing the appropriate connection probability $r(x)$~\cite{Boguna:2004eh}. Hereafter, we use the maximally random ensemble with connection probability given in Eq.~(\ref{r_typeI}). This particular ensemble has, in the thermodynamic limit, only two free parameters, the exponent of the degree distribution $\gamma$ and the average degree $\langle k \rangle$. Notice that $\kappa_0$ is a dummy parameter that can be absorbed in the definition of the hidden variable $\kappa$ so that it can be set to unity at any moment. However, it is useful to keep it during the transformation rule that we apply below. Unlike the regular configuration model (where the actual degrees are fixed a priori) nodes in the canonical configuration model can end up having zero degree and, therefore, the average degree $\langle k \rangle$ can take any positive value, even below 1. 

As already discussed, ensemble self-similarity is always tied to a particular prescription to extract subgraphs out of a given graph. In the case of ensembles of scale-free networks, the natural transformation rule selects subgraphs by removing all nodes with degrees lower than a given threshold value. In the case of the SCM, the transformation rule $T$ removes nodes with hidden variable $\kappa$ below an arbitrary threshold $\kappa_T> \kappa_0$. In~\cite{mariangels,Serrano:2011kq}, we proved that the ensemble of subgraphs so obtained is the same as the original one but with a transformed average degree
\begin{equation}
\langle k \rangle_T=\langle k \rangle \left(\frac{N}{N_T} \right)^{\frac{3-\gamma}{\gamma-1}}  =\langle k \rangle\left(\frac{\kappa_T}{\kappa_0}\right)^{3-\gamma} .
\end{equation}
This simple result provides important insights on how hubs are organized within the network. We first notice that by varying continuously the threshold $\kappa_T$, we obtain a nested sequence of subgraphs.
When $\gamma>3$, $\langle k \rangle_T$ is a monotonic decreasing function of $\kappa_T$. This implies that subgraphs made of high degree nodes are very sparsely connected among them. Thus, even if the original graph is globally connected, connectivity between two hubs is always mediated by chains of low degree nodes. When $\gamma<3$, $\langle k \rangle_T$ is a monotonic increasing function of $\kappa_T$. In turn, this implies that, in the thermodynamic limit, any graph always contains subgraphs made of hubs with arbitrary high connectivity, even if the average degree of the original graph $\langle k \rangle$ is arbitrarily small. This implies that such graphs always have a giant connected component and, so, the original network has a zero percolation threshold~\cite{Serrano:2011kq}.

\subsection{Generalization of the soft configuration model for multiplexes: Self-similarity properties} 

In this paper, we restrict our analysis to self-similar multiplexes with two layers. Generalizations to more than two layers or other ensembles is again straightforward. In the two-layered SCM, each node is characterized by two hidden variables, $\kappa_a$ and $\kappa_b$, distributed according to 
\begin{equation}
\rho(\kappa_a,\kappa_b)=\frac{1}{\kappa_{a0} \kappa_{b0}} \hat{\rho}\left( \frac{\kappa_a}{\kappa_{a0}},\frac{\kappa_b}{\kappa_{b0}}\right),
\end{equation}
with $\kappa_a\ge \kappa_{a0}$, $\kappa_b\ge \kappa_{b0}$, and $\int_1^{\infty} \int_1^{\infty} \hat{\rho}(x,y)dx dy=1$. In this way, $\langle \kappa_{a} \rangle$ and $\langle \kappa_{b} \rangle$ are proportional to parameters $\kappa_{a0}$ and $\kappa_{b0}$ so that they can be set to unity at any moment. In each layer, pairs of nodes connect with connection probabilities 
$r_a(\mu_a \kappa_a \kappa'_a)$ and $r_b(\mu_b \kappa_b \kappa'_b)$, where parameters $\mu_a$ and $\mu_b$ read
\begin{equation}
\mu_a=\frac{\langle k_a \rangle}{N r'_a(0) \langle \kappa_a \rangle^2} \mbox{ and }
\mu_b=\frac{\langle k_b \rangle}{N r'_b(0) \langle \kappa_b \rangle^2}.
\label{mus}
\end{equation}
Notice that the only relation between the two layers comes from the joint distribution $\rho(\kappa_a,\kappa_b)$, which may encode interlayer degree-correlations. 

As for the transformation rule $T$, analogously to the case of single networks, given a multiplex generated from this ensemble, we remove nodes in the multiplex such that their hidden variables $\kappa_a$ and $\kappa_b$ in each layer are below certain threshold values $\kappa_{aT}$ and $\kappa_{bT}$. Next, we analyze under which conditions the multiplex SCM is self-similar.

\subsubsection{Self-similar scale-free multiplexes with uncorrelated interlayer degrees}
When $\kappa_a$ and $\kappa_b$ are uncorrelated variables, the joint degree distribution corresponds to the factorization of the degree distributions of each layer, so that self-similar ensembles of subgraphs can only be achieved if the one-layer degree distributions are scale-free, that is,
\begin{equation}
\hat{\rho}(x,y)=(\gamma_a-1)(\gamma_b-1)/x^{\gamma_a} y^{\gamma_b}.
\label{rhouncorr}
\end{equation}
Thus $\hat{\rho}(x,y)$ is the factorization of two homogeneous functions of degrees $-\gamma_a$ and $-\gamma_b$, which gives a bi-dimensional homogeneous function of degree $-\alpha=-(\gamma_a+\gamma_b)$. After the transformation, the remaining nodes in the subgraph are distributed according to the same scale-free distributions once we replace $\kappa_{a0} \rightarrow \kappa_{aT}$ and $\kappa_{b0}\rightarrow \kappa_{bT}$. The number of nodes that remain in the subgraph is 
\begin{equation}
N_T=\left(\frac{\kappa_{a0}}{\kappa_{aT}}\right)^{\gamma_a-1}\left(\frac{\kappa_{b0}}{\kappa_{bT}}\right)^{\gamma_b-1} N.
\label{N_T}
\end{equation}
The transformation does not change neither the hidden variables of filtered nodes nor their connection probability, which implies that parameters $\mu_a$ and $\mu_b$ remain invariant in the subgraph. Therefore, by combining Eq.~(\ref{mus}) and Eq.~(\ref{N_T}), we conclude that the transformed ensemble is self-similar with re-scaled average degrees
\begin{equation}
\langle k_a \rangle_T=\left(\frac{\kappa_{b0}}{\kappa_{bT}}\right)^{\gamma_b-1} \left(\frac{\kappa_{aT}}{\kappa_{a0}}\right)^{3-\gamma_a} \langle k_a \rangle
\label{average_a}
\end{equation}
and
\begin{equation}
\langle k_b \rangle_T=\left(\frac{\kappa_{a0}}{\kappa_{aT}}\right)^{\gamma_a-1} \left(\frac{\kappa_{bT}}{\kappa_{b0}}\right)^{3-\gamma_b} \langle k_b \rangle.
\label{average_b}
\end{equation}
Notice that in multiplexes with uncorrelated degrees the two thresholds, $\kappa_{aT}$ and $\kappa_{bT}$, are completely independent.

\subsubsection{Self-similar scale-free multiplexes with correlated degrees}
In multiplexes with correlated degrees, self-similarity is achieved when the joint distribution $\hat{\rho}(x,y)$ is a bi-dimensional homogeneous function of degree $\alpha$, that is,
\begin{equation}
\hat{\rho}(a x,a y)=a^{-\alpha} \hat{\rho}(x,y) \; \; \; \forall a.
\label{eq:homogeneous}
\end{equation}
When the degrees in each layer are correlated, this condition enforces a relation between the two thresholds, i.~e. $\kappa_{aT}/\kappa_{a0}=\kappa_{bT}/\kappa_{b0}$, which are not independent anymore~\footnote{For the two thresholds to be independent and the ensemble self-similar, one would need a scaling relation of the type $\hat{\rho}(ax,by)=a^{-\alpha}b^{-\beta} \hat{\rho}(x,y)$. However, the only function in $\mathbb{R}^2$ that satisfy this condition is the factorization of two power laws, which correspond to the case of a multiplex without degree correlations.}. Using the homogeneity property Eq.~\eqref{eq:homogeneous}, it is easy to check that the number of nodes within a subgraph with $\kappa_a>\kappa_{aT}$ and simultaneously $\kappa_b>\kappa_{bT}=\kappa_{b0} \kappa_{aT}/\kappa_{a0}$ is
\begin{equation}
N_T=\left( \frac{\kappa_{a0}}{\kappa_{aT}}\right)^{\alpha-2}N.
\label{N_T2} 
\end{equation}
As in the case of uncorrelated multiplexes, the transformation does not change neither the hidden variables of filtered nodes nor their connection probability, which implies that parameters $\mu_a$ and $\mu_b$ remain invariant in the subgraph. Then, by combining Eq.~(\ref{mus}) and Eq.~(\ref{N_T2}) we conclude that the ensemble is self-similar with re-scaled average degrees in each layer
\begin{equation}
\langle k_a \rangle_T=\left[\frac{\kappa_{aT}}{\kappa_{a0}}\right]^{4-\alpha}\! \! \!  \langle k_a \rangle
\;\mbox{ and }\;
\langle k_b \rangle_T=\left[\frac{\kappa_{aT}}{\kappa_{a0}}\right]^{4-\alpha}\! \! \! \langle k_b \rangle.
\label{average_degrees_T}
\end{equation}

\section{Stability of mutually percolated states in self-similar scale-free multiplexes}
\label{sec:IV}
As mentioned in the introduction, the stability properties of systems of networks can be radically different as compared to single networks depending on the patterns of connectivity between layers~\cite{BPPSH10,Buldyrev:2011fk,Reis:2014}. We shall show that self-similarity can explain several of the previous results on the robustness of systems of networks and can predict new behaviors in a large class of self-similar multiplexes. Notice that the results presented here are qualitatively valid in multiplex ensembles beyond the SCM if those present similar self-similarity properties. 

We study stability in terms of the robustness of the percolated state. In multiplexes, the percolated state can be defined according to different criteria. Here, we assume that nodes in each layer mutually depend on nodes in other layers and that only the nodes that belong to the giant mutually connected component remain functional. The giant mutually connected component of a multiplex network (MCC) is defined as the largest set of nodes that are mutually connected by at least one path in each layer traversing nodes in the MCC~\cite{BPPSH10,Son:2012}. 

For single networks, perturbations in the form of a random failure of a fraction of $1-p$ nodes produce typically a critical phase transition for a specific value $p_c$, so that below $p_c$ the network is fragmented into small components. In multiplexes with a MCC, perturbations can propagate back and forth between the layers so that even small initial failures can produce avalanches of damage leading to a discontinuous collapse of the MCC~\cite{BPPSH10}. Site percolation on random multiplexes has shown indeed a discontinuous hybrid transition at some finite value of the number of nodes removed, where the size of the MCC drops abruptly to zero, like in a first order transition, while the critical behavior is only observed above the transition, like in a second-order one~\cite{BPPSH10,Baxter:2012}. So, perturbations are amplified by the interaction between the layers and systems of networks are said to be more fragile as compared to single networks. The presence of interlayer degree correlations can however revert the situation~\cite{Parshani:2010fk}. Interdependent networks with mutually dependent nodes having identical degrees are statistically more robust than random coupled networks with the same degree distribution. Besides, when $\gamma < 3$, they disintegrate via a second-order phase transition ---in the same way as noninteracting networks--- and are thus very resilient against random failures~\cite{Buldyrev:2011fk}. More structured systems of correlated interconnected networks or with overlaps have been proved to be robust to failure as well~\cite{Cellai:2013fk,Reis:2014}.

Next, we assess the resilience of MCCs to random failures in scale-free multiplexes on the basis of their self-similarity properties and check numerically our predictions. Before that, we note that the average degree $\langle k \rangle$ in the SCM ensemble defined in Sec.~\ref{sec:III} is equivalent to the site percolation probability $p$ and it can then be used in robustness studies as the control parameter. Indeed, when a random fraction of $1-p$ nodes is removed from a given graph of the ensemble, the hidden variables $\kappa$s of the remaining nodes are distributed as in the original graph and the connection probability among them remains unchanged. However, the number of nodes in the subgraph is $pN$. Since $\mu$ remains unchanged, Eq.~(\ref{mu}) implies that this ensemble is self-similar under a random removal of nodes with a modified average degree $\langle k \rangle_T=p\langle k \rangle$. This means that, in the thermodynamic limit, removing a random fraction of nodes $1-p$ of a network with average degree $\langle k \rangle$ is equivalent to generating a graph of the same ensemble but with an average degree $p \langle k \rangle$. Because of this equivalence, hereafter we use $\langle k \rangle$ as the control parameter of the percolation properties of the ensemble. 

\subsection{Fragility of uncorrelated scale-free multiplexes explained by self-similarity}
Single scale-free self-similar networks in the thermodynamic limit with $\gamma<3$ always contain subgraphs made of hubs with arbitrary high connectivity, even if the average degree of the original graph $\langle k \rangle$ is arbitrarily small, which implies that such graphs always have a giant connected component and, so, a zero percolation threshold~\cite{Serrano:2011kq}. This makes such structures robust to random failures. In the case of uncorrelated multiplexes, the question is whether it is still possible to find a continuous set of nested subgraphs such that the average degrees within the subgraphs increase in both layers simultaneously. In that case the multiplex would be robust to random failures, being able to maintain a MCC despite perturbations.
\begin{figure}[t]
\centerline{\includegraphics[width=0.8\linewidth]{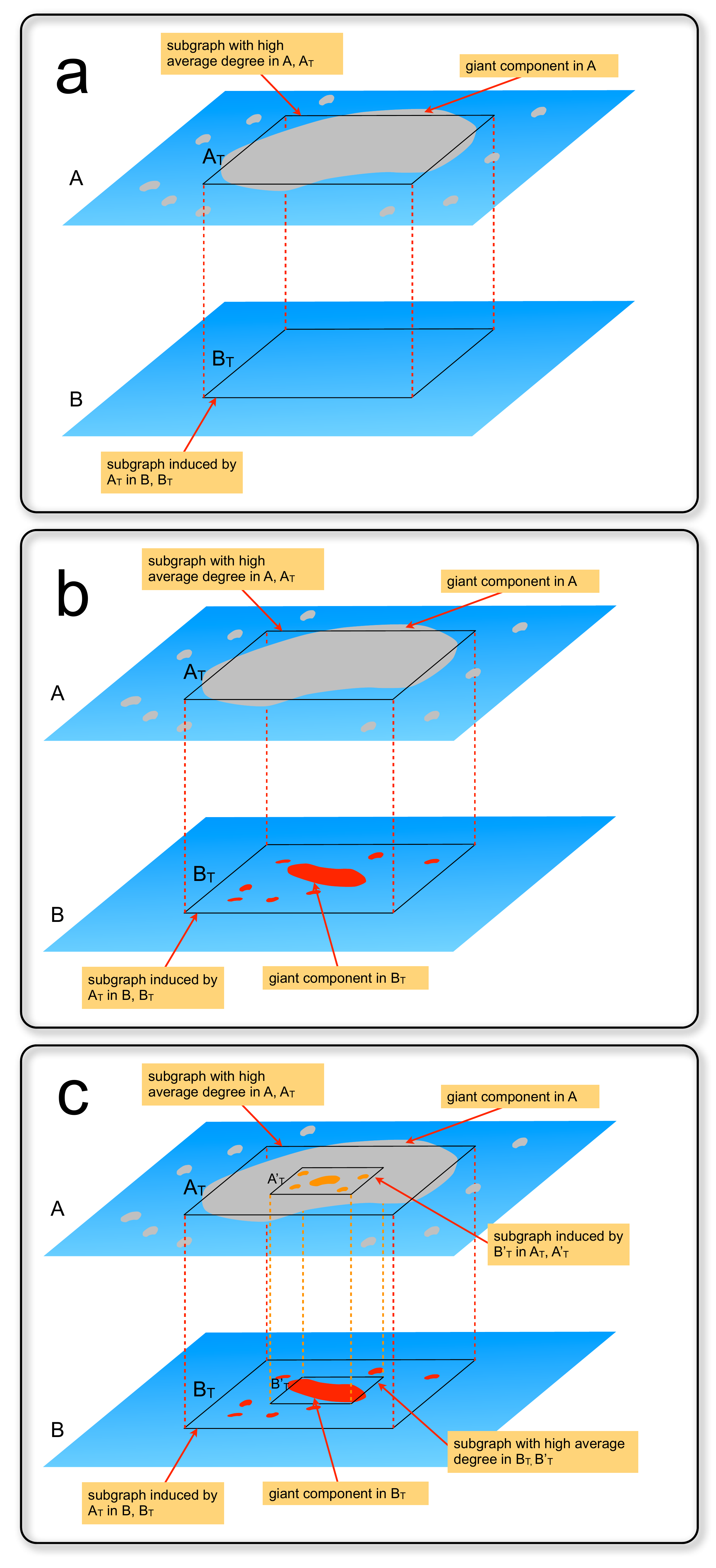}}
\caption{Sketch of a self-similar multiplex with two layers, A and B, with uncorrelated degrees across layers. In panel {\bf a}, A$_T$ is A's subgraph made of nodes with high degrees, such that the giant component of layer A is almost contained in A$_T$. The set of nodes in A$_T$ induces a subgraph in layer B, B$_T$. In panel {\bf b}, because of the self-similarity properties of the ensemble, B$_T$ is similar to B but, due to the impossibility to satisfy simultaneously the inequalities in Eq.~\eqref{inequality}, it has a smaller average degree than B. Thus, its giant component ---which is the potential candidate to be the MCC of the multiplex--- is also reduced. In panel {\bf c}, this process can be iterated {\it at infinitum} and can lead to the fragmentation of the multiplex. 
\label{fig:2}} 
\end{figure}

To have a nested ensemble of subgraphs, $\kappa_{bT}$ must be either constant or a monotonic increasing function of $\kappa_{aT}$ (or vice versa). Let $\kappa_{bT}=g(\kappa_{aT})$ be such function. Then, the condition for Eqs.~(\ref{average_a}) and (\ref{average_b}) to be simultaneously monotonic increasing functions of $\kappa_{aT}$ is
\begin{equation}
\frac{\gamma_a-1}{3-\gamma_b}<\frac{\kappa_{aT}g'(\kappa_{aT})}{\kappa_{a0}g(\kappa_{aT})}<\frac{3-\gamma_a}{\gamma_b-1}.
\label{inequality}
\end{equation}
However, these inequalities can only hold if the lower bound is smaller than the upper bound, which is equivalent to the inequality $\alpha=\gamma_a+\gamma_b<4$. This is clearly not possible in scale-free sparse graphs with $\gamma_a$ and $\gamma_b$ in the range $(2,3)$, implying that, while it is possible to have a sequence of subgraphs with increasing average degree in one of the layers (if one of the inequalities is satisfied), the same sequence of subgraphs has necessarily a decreasing average degree in the other layer. 

This result explains the fragility of scale-free systems of networks first reported in~\cite{BPPSH10}. In single scale-free networks, global connectivity is mainly provided by the interconnection of high degree nodes, which gives the main explanation for their robustness. In uncorrelated scale-free multiplexes, the situation is different. Our self-similarity argument starts by selecting a subgraph of high degree nodes in layer A and so an almost fully connected subgraph that contains the majority of nodes of the giant component of layer A (see panel {\bf a} in Fig.~\ref{fig:2}). However, as our previous result shows, the average degree in layer B of the subgraph induced by the subgraph in A is smaller than in B and, thus, its giant component in B ---which is the candidate set to contain the MCC of the mutually percolated multiplex--- is also reduced. We could now select a subgraph of the subgraph in layer B such that its average degree is high enough to contain its layer B giant component. However, the average degree of the induced sub-subgraph in layer A will decrease below its original value, and so its giant component. This process can be iterated {\it at infinitum} and, at each iteration, the size of the potential subgraph to contain a MCC is reduced. We thus conclude that the MCC cannot be sustained by high degree nodes alone and must rely on the connectivity of low degree nodes. This makes scale-free multiplexes always more fragile than more homogeneous networks with the same average degree. 
\begin{figure}[t]
\centerline{\includegraphics[width=\linewidth]{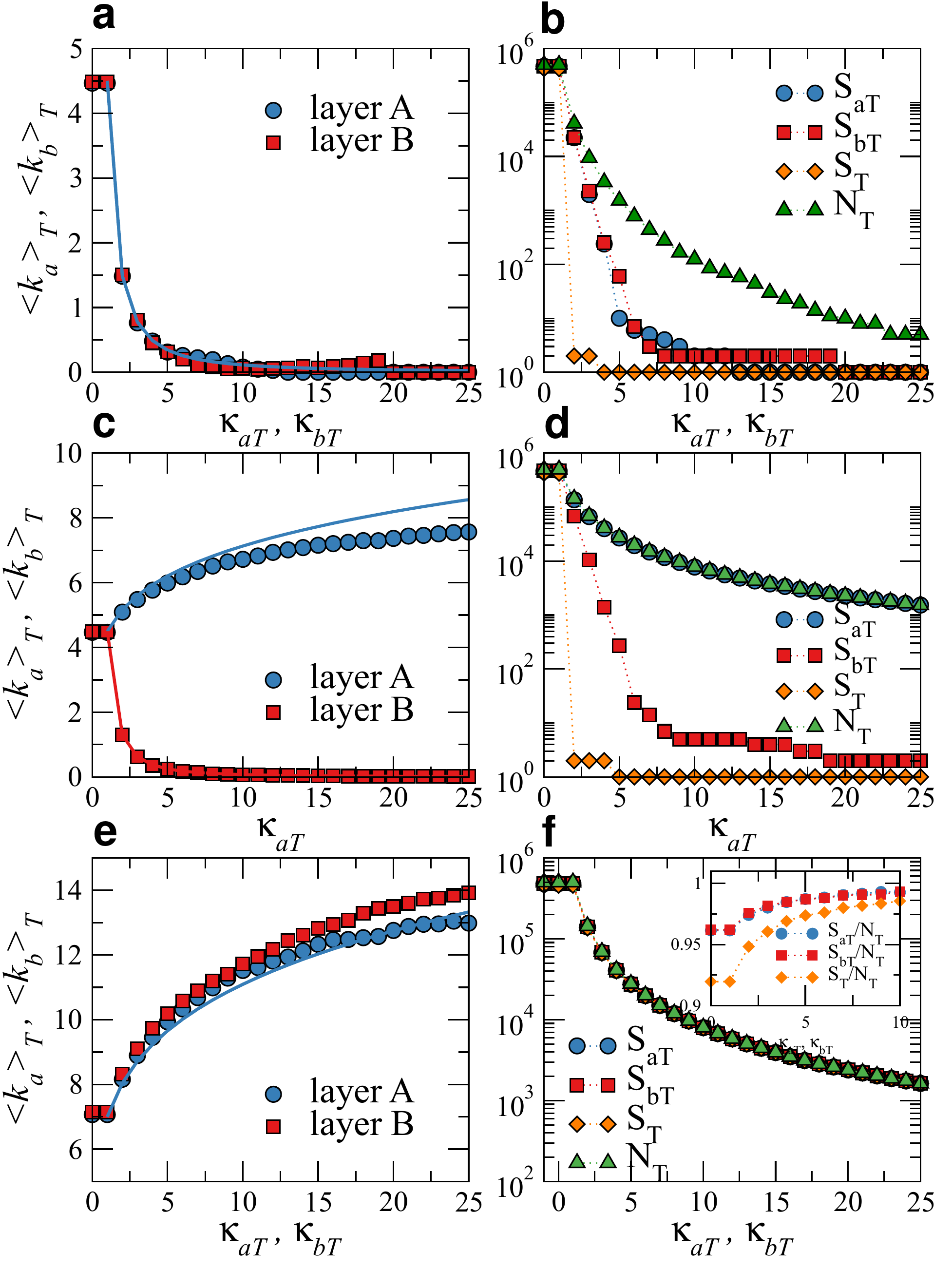}}
\caption{Average degrees (left column) and the size of the largest connected components (right column) as a function of filtering parameters $\kappa_{aT}$ and $\kappa_{bT}$. Panels {\bf a} and {\bf b} show results for a multiplex network with uncorrelated degrees where $\kappa_{aT} = \kappa_{bT}$. Panels {\bf c} and {\bf d} show results for a multiplex network with uncorrelated degrees where $\kappa_{bT}$ is fixed to the minimum value $\kappa_{b0}$. Panels {\bf e} and {\bf f} show the results for a multiplex network with correlated degrees where $\kappa_{aT} = \kappa_{bT}$. Solid lines correspond to the analytical results given by Eqs.~\eqref{average_a}, \eqref{average_b} and \eqref{average_degrees_T}. In all cases, the multiplex network is composed of two layers with $N =5 \times 10^5$ nodes, $\gamma = 2.8$, $\langle k_{min} \rangle = 2$, and we evaluated the absolute size of the largest connected components $S_{aT}$ and $S_{bT}$ in individual layers $A$ and $B$, the size of the MCC $S_{T}$, and the size of the network $N_{T}$ after applying the corresponding transformation.\label{fig:3}} 
\end{figure}

\subsection{Robustness of correlated scale-free multiplexes explained by self-similarity}
The picture changes completely when the degrees in each layer are positively correlated. 
In the case of sparse scale-free self-similar multiplexes with uncorrelated degrees in the two layers, $\alpha=\gamma_a+\gamma_b>4$ so that the conditions for a stable MCC are not fulfilled. However, when $\kappa_a$ and $\kappa_b$ are positively correlated, it is possible to find ensembles with $3<\alpha<4$. As an example, consider the joint distribution
\begin{equation}
\hat{\rho}(x,y)=\frac{\gamma (\gamma-1)2^{\gamma-1}}{(x+y)^{\gamma+1}}.
\label{rhocorr}
\end{equation}
Its marginal distribution is $\hat{\rho}(x)=(\gamma-1) 2^{\gamma-1}(1+x)^{-\gamma}$~\footnote{Notice that a homogeneous distribution in two dimensions does not imply that its marginal is also a homogeneous function.}.  From here, the conditional average is $\langle x | y \rangle=(y+\gamma)/(\gamma-1)$, so that the correlation between $x$ and $y$ increases when $\gamma \rightarrow 2$. The joint distribution Eq.~\eqref{rhocorr} is a homogeneous function with $\alpha=\gamma+1$. Therefore, according to Eqs.~\eqref{average_degrees_T}, when $\gamma<3$ the ensemble has self-similar subgraphs with increasing average degree in both layers simultaneously. This readily implies that the ensemble always possesses a MCC so that its percolation threshold is zero in the thermodynamic limit. Besides, the ``transition'' is continuous, in the sense that the relative size of the MCC approaches zero monotonously when $p \rightarrow 0$. This generalizes the result found in~\cite{Buldyrev:2011fk} for networks with identical degrees in both layers and makes an important step forward as it quantifies the precise level of correlations (and so the value of $\alpha$) that is needed to go from a hybrid discontinuous transition to a continuous one.

\subsection{Numerical simulations}

To check numerically the predicted stability properties of self-similar scale-free multiplexes, we generated two-layered multiplexes using the canonical configuration model. In all cases, $N=5\times 10^5$ and $\langle k_{min} \rangle = 2$. For uncorrelated scale-free multiplexes, we used the joint probability distribution Eq.~(\ref{rhouncorr}), while we implemented correlations according to Eq.~(\ref{rhocorr}). Finally, to compute mutually connected components, we implemented an efficient algorithm based on Ref.~\cite{Hwang:2014}, which keeps track of all the MCCs, not only the giant, present in a multiplex. The algorithm represents each layer of the multiplex by the dynamic connectivity structure defined in~\cite{Holm:2001fk}. This structure allows for maintaining information about network components and their sizes, while updating a graph by deletion or insertion of edges. The algorithm works in two phases. First, we find MCCs in the initial multiplex and second, we calculate the size of the giant MCC for all values of the parameter $p$. 

To compute all MCCs in the initial multiplex, we identify connected components for each layer separately and if needed, we reconnect all single components by adding a minimum number of ad hoc edges. Thus, after this step every layer is a single connected component and the multiplex a single MCC. Next, we sequentially delete all add hoc edges. Each single removal creates two separated components in the given layer. We then check all possible node pairs, where each node in the pair belongs to a different component and remove, in all other layers, edges connecting them. Whenever any removed edge breaks a connected component into two, we have to continue with the removal of all edges that connect disconnected components in all other layers. Finally, when all ad hoc edges are removed, all layers consist of connected components corresponding to MCCs. 
In the second phase, we generate a random sequence defining the order of node removals. Removal of each node is accomplished by removing all its adjacent edges from all layers. Every edge is removed in the same way as ad hoc edges in the first phase of the algorithm. Similarly as in the first phase, after removing the node all layers consist of connected components corresponding to MCCs. The size of the largest component is outputted as the size of the largest MCC for the corresponding $p$ value.

In Fig.~\ref{fig:3}, we show the average degrees in the subgraphs and the size of the largest connected components in each layer and the MCC as a function of the filtering thresholds $\kappa_{aT}$ and $\kappa_{bT}$. In all cases, networks are scale-free with $\gamma=2.8$. In uncorrelated multiplexes, the average degrees of the subgraphs cannot increase simultaneously as the thresholds increase. This is shown in Fig.~\ref{fig:3}~a for $\kappa_{aT} = \kappa_{bT}$ and in Fig.~\ref{fig:3}~c for $\kappa_{bT} = \kappa_{b0}=1$. As clearly seen in the figures, the only possibilities are that the average degrees decrease simultaneously (when $\kappa_{aT}=\kappa_{bT}$) or that the average degree of one of the layers increases while the other decreases (when $\kappa_{bT}$ is constant). This induces the fragility of the MCC which, as shown in Fig.~\ref{fig:3}~b and d, reduces its size abruptly at some relatively small value of the threshold. Interlayer degree correlations change completely the picture. In Fig.~\ref{fig:3}~e and f, we show the average degrees in the subgraphs and the size of the different components for $\kappa_{aT} = \kappa_{bT}$ in a canonical configuration model multiplex ensemble with the joint degree distribution given by Eq.~(\ref{rhocorr}). In this case, it is possible to produce sequences of subgraphs with increasing average degrees in both layers simultaneously, so that the MCC becomes very robust. Finally, the inset in Fig.~\ref{fig:3}~f shows the relative size of the MCC (relative to the remaining number of nodes after the filtering process), which approaches 1 for large values of the thresholds, indicating that, as predicted, such self-similar multiplex contains a small but macroscopic subgraph that is completely connected in both layers simultaneously.

\begin{figure}[t]
\centerline{\includegraphics[width=\linewidth]{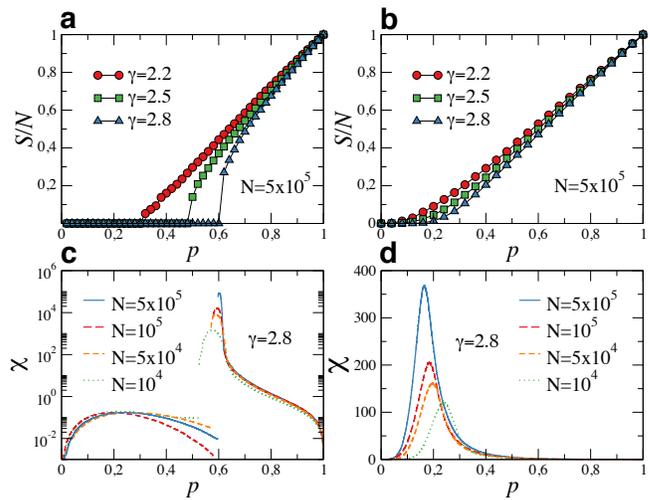}}
\caption{Comparison between the percolation properties of scale-free multiplexes with uncorrelated (left panel) and correlated (right panel) degrees. Panels {\bf a} and {\bf b} show the relative size of the largest mutually connected component vs the fraction $p$ of nodes remaining undamaged. In both cases $N = 5 \times 10^5$ and $\gamma = 2.2$, $2.5$, and $2.8$. Each curve corresponds to one complete random sequence of node removals. Panels {\bf c} and {\bf d} show the susceptibility $\chi$ as a function of site occupation probability $p$ for scale free multiplexes of $\gamma=2.8$ and different sizes. The different curves $\chi(p)$ are computed from $10^4$ complete random sequences of node removals. In all cases, multiplexes are composed of two layers and $\langle k_{min} \rangle = 2$.\label{fig:4}} 
\end{figure}

To get further insights into the percolation properties of self-similar multiplexes, we adopt the conventional percolation criterion of measuring the breakdown of the largest MCC. We computed the relative size of the largest MCC versus the fraction of nodes $p$ remaining in the multiplex for different values of the power-law exponent $\gamma$. Results are shown in Fig.~\ref{fig:4}~a for multiplexes with uncorrelated degrees and in Fig.~\ref{fig:4}~b for correlated ones. For all values of $\gamma$, the transition between the mutually percolated and the fragmented states is discontinuous in the uncorrelated case while it is continuous and approaching zero in the correlated case. This can be corroborated by the scaling of the susceptibility vs the system size, where the susceptibility $\chi$ is defined as
\begin{equation}
\chi=\frac{\langle S^2 \rangle-\langle S \rangle^2}{\langle S \rangle}.
\end{equation}
Here $S$ is the size of the largest MCC at any value of $p$ and averages are taken over a large number of complete random sequences of node removals. This quantity is able to distinguish between discontinuous, continuous, and hybrid phase transitions. In continuous phase transitions, $\chi$ shows a clear peak close to the critical point that diverges as the system size increases. Instead, in discontinuous transitions, $\chi$ shows a discontinuity at the critical point but no dependence on the system size. In the case of hybrid phase transitions, $\chi$ shows a diverging peak approaching the critical point from one side, a discontinuity and then a size independent behavior on the other side. According to these criteria, Fig.~\ref{fig:4}~c indicates that the transition is hybrid in multiplexes with uncorrelated degrees whereas Fig.~\ref{fig:4}~d indicates that $\chi$ has a continuous divergence with a peak that approaches zero in the thermodynamic limit. This clearly corroborates our theoretical prediction about a zero percolation threshold but with critical fluctuations when $p \rightarrow 0$.

\section{Conclusions}
\label{sec:V}
Self-similarity is a widespread property in network models and has also been observed in many real world networks~\cite{mariangels}. Beyond the mathematical beauty of self-similarity, this property has important implications for the structural properties of networks. The power of the concept was illustrated in single-layered networks by the proof of a zero percolation threshold for a general class of self-similar networks, which only required the self-similarity property with a hierarchy of nested subgraphs whose average degrees grow with their depth in the hierarchy~\cite{Serrano:2011kq} and without the need of usual limiting requirements. 

In this paper, we have extended the concept to multiplexes and illustrated its importance by assessing the stability of scale-free multiplexes in terms of their self-similarity properties. To state in a clear and explicit way the definition and relevance of self-similarity, we have focused on the SCM ensemble. However, we should stress that the results presented here are qualitatively valid in other multiplex ensembles with similar features, that is, with similar self-similarity properties, degree distributions and interlayer degree correlations. Interestingly, the observed fragility of scale-free multiplexes or the robustness to failure of correlated systems of networks can be explained and predicted based only on their self-similarity characteristics. In particular, we have found that scale-free multiplexes can recover a zero percolation threshold and a continuous transition in the thermodynamic limit, and so the ordinary stability properties of single scale-free networks. Self-similarity can as well have important implications for other critical phenomena taking place in multiplex structures when the critical point is a function of the connectivity of the system.

\begin{acknowledgments}
This work was supported by a James S. McDonnell Foundation Scholar Award in Complex Systems; the European Commission LASAGNE project no.\ 318132 (STREP); the ICREA Academia prize, funded by the {\it Generalitat de Catalunya}; the MINECO project no.\ FIS2013-47282-C2-1-P; the {\it Generalitat de Catalunya} grant no.\ 2014SGR608; and  {\it APVV} (project APVV-0760-11).
\end{acknowledgments}

\end{document}